\documentclass[num-refs]{wiley-article}

\usepackage{siunitx}
\usepackage{multirow}
\usepackage{booktabs}
\usepackage{lscape}

\newcommand\independent{\protect\mathpalette{\protect\independenT}{\perp}}
\def\independenT#1#2{\mathrel{\rlap{$#1#2$}\mkern2mu{#1#2}}}

\DeclareMathOperator{\E}{\textnormal{\mbox{E}}}

\papertype{Tutorial}

\title{Extending inferences from a randomized trial to a new target population}

\abbrevs{DR, doubly robust; IO, inverse odds; OM, outcome model-based.}

\author[1-4]{Issa J. Dahabreh MD ScD}
\author[1,2]{Sarah E. Robertson MS}
\author[5]{Jon A. Steingrimsson PhD}
\author[6]{Elizabeth A. Stuart PhD}
\author[4,7,8]{Miguel A. Hern\'an MD DrPH}

\affil[1]{Center for Evidence Synthesis in Health, Brown University, Providence, RI}
\affil[2]{Department of Health Services, Policy \& Practice, Brown University, Providence, RI}
\affil[3]{Department of Epidemiology, Brown University, Providence, RI}
\affil[4]{Department of Epidemiology, Harvard T.H. Chan School of Public Health, Boston, MA}
\affil[5]{Department of Biostatistics, School of Public Health, Brown University, Providence, RI}
\affil[6]{Departments of Mental Health, Biostatistics, and Health Policy and Management, Johns Hopkins Bloomberg School of Public Health, Baltimore, MD}
\affil[7]{Department of Biostatistics, Harvard T.H. Chan School of Public Health, Boston, MA}
\affil[8]{Harvard-MIT Division of Health Sciences and Technology, Boston, MA}

\corraddress{Issa Dahabreh MD ScD, Box G-S121-8; Brown University, Providence, RI 02912.}
\corremail{issa\_dahabreh@brown.edu}

\fundinginfo{PCORI, Methods Research Awards Number: ME-1306-03758, ME-1502-27794, ME-1503-28119.}

\runningauthor{Issa Dahabreh et al.}

\begin{document}

\maketitle

\begin{abstract}
When treatment effect modifiers influence the decision to participate in a randomized trial, the average treatment effect in the population represented by the randomized individuals will differ from the effect in other populations. In this tutorial, we consider methods for extending causal inferences about time-fixed treatments from a trial to a new target population of non-participants, using data from a completed randomized trial and baseline covariate data from a sample from the target population. We examine methods based on modeling the expectation of the outcome, the probability of participation, or both (doubly robust). We compare the methods in a simulation study and show how they can be implemented in software. We apply the methods to a randomized trial nested within a cohort of trial-eligible patients to compare coronary artery surgery plus medical therapy versus medical therapy alone for patients with chronic coronary artery disease. We conclude by discussing issues that arise when using the methods in applied analyses.

\end{abstract}

\section{Background}

Participants in randomized trials often differ from individuals seen in clinical practice with respect to the distribution of covariates that modify the treatment effect. Because of such differences, the average treatment effect in the population of trial participants differs from the average treatment effect in other populations \cite{rothwell2005}. This lack of generalizability or transportability of trial results has been documented by numerous surveys comparing the distribution of baseline covariates between trial participants and samples of patients seen in clinical practice (e.g., \cite{Evans2001, Elting2006, Steg2007}).

The task of assessing whether trial findings are relevant to a new target population is often handled via informal subjective assessments \cite{dans1998users}. Clinicians and policy makers, however, cannot easily use such assessments to draw inferences about the target population they are interested in (e.g., the population of patients in their practice or the population to which the policy will be applied) and they cannot evaluate the performance of informal assessments for guiding decisions. 

This tutorial builds on a growing literature (e.g., \cite{cole2010, kaizar2011, omuircheartaigh2014, tipton2012, tipton2014, hartman2013, zhang2015, rudolph2017, westreich2017, buchanan2018generalizing, dahabreh2018generalizing}) to show that, instead of relying exclusively on informal assessments, we can use formal statistical methods to extend causal inferences from trial participants to a new target population. We focus on the common setting in which individual-level data on time-fixed treatments, outcomes, and baseline covariates are available from a randomized trial, but only individual-level data on baseline covariates are available from the target population of non-participants. We examine methods for extending inferences using models for the expectation of the outcome, the probability of trial participation, or both (doubly robust). We assess the finite-sample performance of the methods in simulation studies and provide example code to implement the methods in \texttt{R}. We apply the methods to a randomized trial nested within a cohort of eligible patients with coronary artery disease that compared coronary artery surgery plus medical therapy versus medical therapy alone. Last, we discuss issues that arise when using the methods in practice.

\section{Extending trial findings to a target population}\label{section_designs}

\subsection{Basic concepts and terminology}

Suppose that we have access to data from a completed trial that compared two or more treatments. We want to evaluate the comparative effectiveness of the treatments studied in the trial in some well-defined \emph{target population of non-participants}. Usually, the distribution of effect modifiers differs between the population represented by trial participants and the target population. Such differences are due to the trial participation process \cite{dahabreh2019commentaryonweiss}: First, the investigators specify eligibility criteria, which define the \emph{trial-eligible population}. Next, they invite eligible individuals to participate in the trial and the recruitment activities define the \emph{invited population}. Last, the invited individuals who chose to participate in the trial define the \emph{participant population}. 

We use the term \emph{generalizability} when the target population coincides or is a subset of the trial-eligible population and \emph{transportability} when the target population includes at least some individuals who are not trial-eligible (and who, by definition, cannot be trial participants) \cite{dahabreh2019commentaryonweiss, hernan2016discussionkeiding} (others have proposed different definitions \cite{westreich2017}). In generalizability analyses, the target population often has a different distribution of effect modifiers compared with the participant population \cite{dahabreh2018generalizing}, even though both populations meet the trial eligibility criteria. In transportability analyses, such differences are more likely.

We collectively refer to generalizability and transportability as \emph{extending inferences from a trial to a target population}. The methods described in this paper apply to analyses extending inferences (i.e., both generalizability and transportability) from a trial to a target population of non-participants. Henceforth, for brevity, we mostly use the term ``target population'' without the qualifier ``of non-participants;'' it should be understood, however, that our results are limited to that population.

\subsection{Study designs}

The most natural study design for extending inferences from a randomized trial to the target population is a \emph{nested trial design}, in which the randomized trial is embedded within a simple random sample of individuals from a broader population, including individuals who refuse to participate in the trial and those who are not invited \cite{dahabreh2019studydesigns}. For example, in comprehensive cohort studies \cite{olschewski1985}, a trial is nested within a cohort of trial-eligible individuals. In this design, investigators collect baseline covariate information from all cohort members, but need only collect treatment and outcome information from randomized individuals. This setup also occurs in trials embedded in health-care systems where baseline covariate data are routinely collected from all members of the system, but only a subset of the members participate in the trial \cite{fiore2016}. 

An alternative for extending inferences to the target population is a \emph{non-nested trial design}, in which the randomized trial is combined with a separately obtained sample from the target population \cite{dahabreh2019studydesigns}. Here, investigators create a \emph{composite dataset} by appending the data from a completed trial to a separately obtained sample from the target population (e.g., \cite{westreich2017}); the composite dataset facilitates modeling and estimation, as we discuss below. We assume that the sampled non-participants are a simple random sample from the target population. More specifically, we assume that the trial sample and the sample of non-participants are obtained separately; and that all trial participants, but only a subset of non-participants, are sampled (the latter with unknown sampling probability); see Section B of the Appendix for details about the underlying sampling scheme. As in nested trial designs, baseline covariate information is available both from randomized and non-randomized individuals, but treatment and outcome information is only available from randomized individuals. This setup occurs in drug development and regulatory settings, because treatment and outcome data are available only from a small number of trial participants prior to drug approval, but baseline covariate data can be collected from large samples of individuals who would be eligible for participation in the trials and can be identified in routinely collected data (for treatments that are commercially available, trial-eligible new-users \cite{ray2003} of the treatments compared in the trials may be a reasonable choice of non-participants). The setup also occurs in policy research, when randomized trials are conducted in selected samples but target population data are available from administrative databases or surveys. 

The methods we describe in this tutorial can be used both in nested and non-nested trial designs to estimate treatment effects in the target population, using baseline covariate, treatment, and outcome data from the trial and only baseline covariate data from the target population. Treatment and outcome data from the target population, if available, can be used to evaluate assumptions \cite{hartman2013, dahabreh2018generalizing}, but they are not necessary for the types of analyses described in this tutorial. This is an attractive feature if treatment and outcome data from non-participants are unreliable (e.g., due to confounding of the treatment-outcome association or gross measurement error), costly to obtain, or impossible to collect.

\section{Data and causal quantities of interest} \label{section_estimands}

The data generated from the designs described in the previous section consist of independent observations on baseline covariates $X$; assigned treatment $A$; outcome $Y$; and a trial participation indicator $S$ that takes the value 1 for trial participants or 0 for non-participants. In nested trial designs, the data consist of realizations of independent and identically distributed random tuples $(X_i, S_i, S_i \times A_i, S_i \times Y_i)$, $i = 1, \ldots, n$, where $n$ is the total number of observations and we define $n_{\text{\tiny RCT}}=\sum_{i=1}^{n}I(S_i=1)$ and $n_{\text{\tiny obs}}=\sum_{i=1}^{n}I(S_i = 0)$, where $I(\cdot)$ denotes the indicator function. In non-nested trial designs, the data available from the randomized trial are realizations of $(X_i, S_i = 1, A_i, Y_i)$, $i = 1, \ldots, n_{\text{\tiny RCT}}$; the data from the sample of the target population are realizations of $(X_i, S_i = 0)$, $i = 1, \ldots, \widetilde{n}_{\text{\tiny obs}}$ and we define the total number of (sampled) observations that contribute data to the analysis as $\widetilde n = n_{\text{\tiny RCT}} + \widetilde{n}_{\text{\tiny obs}}$. Note that the number of trial participants, $n_{\text{\tiny RCT}}$, is the same as in nested trial designs, because in non-nested trial designs all trial participants are sampled; but  $\widetilde{n}_{\text{\tiny obs}} \neq n_{\text{\tiny obs}}$ and $\widetilde n \neq n$, because only a subset of the non-participants are sampled (with unknown sampling probability). 

In both designs, the data exhibit a special missingness pattern: for trial participants we have data on $(S = 1, X, A, Y),$ but for non-participants we only have data on $(S = 0, X)$. In the main text of the tutorial, we focus on nested trial designs because we think that they will be increasingly used in the future and because our motivating application had such a design; in the Appendix we discuss the sampling properties and give additional results for non-nested trial designs. Table \ref{table_data_structure} shows the observed data structure for the special case of binary treatment in a nested trial design.

\begin{table}[ht]
	\renewcommand{\arraystretch}{1.3}
	\centering
\caption{Data structure for binary treatment, $A$, observed outcome, $Y$, and baseline covariates, $X$, for trial participants ($S = 1$) and non-participants ($S = 0$) in a nested trial design. The observed data include $(S = 1, X, A, Y)$ information from $n_{\text{\tiny RCT}}$ trial participants, of whom $n_0$ are assigned to $A = 0$ and $n_1$ assigned to $A = 1$; and $(S= 0, X)$ information from $n_{\text{\tiny obs}}$ non-participants; $n$ is the total number of observations. Dashes denote missing data.}
\begin{tabular}{|ccccc|}
		\hline
		Unit &  $S$      & $A$ & $Y$ & $X$         \\ \hline\hline
		$1 $  & 1        & 0   & $y_{1} $ & $x_{1}$  \\ 
		$\vdots$ & 1        & $\vdots$                & $\vdots$   & $\vdots$     \\ 
		$n_0 $  &  1        & 0   & $y_{n_0} $     & $x_{n_0}$     \\ \hline\hline 
		$1 + n_0$ & 1        & 1  & $y_{1 + n_0}$     & $x_{1 + n_0}$ \\ 
		$\vdots$ &  $\vdots$      & $\vdots$   & $\vdots$    & $\vdots$    \\ 
		$n_1 + n_0 = n_{\text{\tiny RCT}}$ & 1        & 1 & $y_{n_1 + n_0} $     & $x_{n_1 + n_0}$ \\ \hline\hline 
		$1 + n_{\text{\tiny RCT}}$   & 0        & $-$ & $-$   & $x_{n_1 + n_0 + 1}$ \\ 
		$\vdots$ & $\vdots$  & $\vdots$   & $\vdots$   & $\vdots$    \\ 
		$n_{\text{\tiny obs}} + n_{\text{\tiny RCT}} = n$   &0        & $-$        & $-$  & $x_{n}$     \\ \hline
	\end{tabular}
	\label{table_data_structure}
\end{table}

We use the random variables $Y^a$ to denote the potential (counterfactual) outcomes under intervention to set treatment to $a$, possibly contrary to fact \cite{splawaneyman1990, rubin1974}. We only consider treatments with a finite number of possible levels in this tutorial; extensions to continuous treatments are possible but not pursued here. For any two treatment values, $a, a'$, we are interested in the average treatment effect in the target population, $\E [Y^a - Y^{a'} | S = 0]$. Because the average treatment effect in randomized trials without loss to follow-up, $\E [Y^a - Y^{a'} | S = 1]$ is identifiable under standard assumptions \cite{hernan2020}, it is interesting to compare $\E [Y^a - Y^{a'} | S = 1]$ against $\E [Y^a - Y^{a'} | S = 0]$. Though often assumed equal, these two effects are different when the effect of treatment varies over baseline covariates that are differentially distributed between trial participants and non-participants (i.e., when the distribution of effect modifiers on the additive scale is different between trial participants and non-participants) \cite{Dahabreh2016, dahabreh2017}.

When using nested trial designs, another causal effect, the average effect in the population represented by the cohort in which the trial is nested, $ \E [Y^a - Y^{a'}] $, may also be of interest, and we have discussed its identification and estimation in previous work \cite{dahabreh2018generalizing}. Even if the primary target of inference for nested trial designs is $ \E [Y^a - Y^{a'}]$, investigators are typically also interested in estimating the effect among non-participants, $ \E [Y^a - Y^{a'} | S = 0]$. Comparing the effect among non-participants against the effect among trial participants is often of substantial scientific and policy interest. Importantly, when using non-nested trial designs, $\E [Y^a - Y^{a'}]$ cannot be identified unless additional information about the sampling of individuals from the target population is available (for more details, see the Appendix of this tutorial, and references \cite{dahabreh2019studydesigns} and \cite{dahabreh2019generalizing}). In the absence of such information, $\E [Y^a - Y^{a'}]$ is not identifiable in non-nested trial designs because the data are a ``mixture'' of trial participants and non-participants, with arbitrary mixing proportions (determined by the size of the trial and the sample from the target population). Thus, in the remainder of this tutorial we focus on the identification and estimation of the components $\E [Y^a | S = 0]$ of the average treatment effect in the target population.

\section{Identifiability conditions \& identification}\label{section_identifiability}

\subsection{Identifiability conditions}

We now discuss sufficient conditions for identifying the mean of the potential outcomes under each treatment $a$ in the target population, $\E [Y^a | S = 0] $.

\noindent
\textbf{(1) Consistency of potential outcomes:} The observed outcome for the $i$th individual who received treatment $a$ equals that individual's potential outcome under the same treatment, that is, for every $i$, if $A_i = a$, then $Y_i = Y^{a}_i$. Implicit in this notation is the assumption that trial participation does not affect the outcomes except through treatment assignment (e.g., there are no Hawthorne effects) \cite{dahabreh2019identification}.

\noindent
\textbf{(2) Conditional exchangeability in mean over $A$ in the randomized population:} Among trial participants, the potential outcome mean under treatment $a$ is independent of treatment, conditional on baseline covariates: $ \E [ Y^{a} | X = x, S = 1, A = a ] = \E [ Y^{a} | X =x , S = 1 ]$ for every $x$ with positive joint density $f_{X,S}(x, S = 1)$. We expect  conditional exchangeability in the trial to hold, regardless of whether the randomization was unconditional or conditional on covariates. The mean exchangeability assumption is weaker than the assumption of exchangeability in distribution over $A$, that is, $Y^{a} \independent A| X, S=1$.
 
\noindent
\textbf{(3) Positivity of treatment assignment:} In the trial, the probability of being assigned to each of the treatments being compared, conditional on the covariates needed for exchangeability over $A$ in the randomized population, is positive: $\Pr[A=a | X = x, S=1] > 0 $ for every $a$ and every $x$ with positive joint density $f_{X,S}(x, S =1)>0$. 
 
\noindent
\textbf{(4) Conditional exchangeability in mean over $S$:} The potential outcome mean is independent of trial participation, conditional on baseline covariates: $\E [ Y^{a} | X = x, S = 1 ] = \E [ Y^{a} | X = x, S = 0 ]$ for every $x$ with positive joint density  $f_{X,S}(x, S = 0)$.  Again, this assumption is weaker than the assumption of exchangeability in distribution over $S$, that is, $Y^{a} \independent S|X$. 
 
\noindent
\textbf{(5) Positivity of trial participation:} The probability of participating in the trial, conditional on the covariates needed to ensure exchangeability over $S$, is positive, $\Pr[S=1 | X = x] >0$, for every $x$ with positive joint density $f_{X,S}(x, S = 0) > 0$. Informally, this condition ensures that, for the covariates that are necessary to achieve exchangeability over $S$, all covariate patterns in the target population will be represented in the trial as the trial sample size gets arbitrarily large. Note that the probability of trial participation does not need to be bounded away from 1, since it is not possible to have $f_{X,S}(x, S = 0) > 0$ and $\Pr[S=1 | X = x] = 1$ for any $x$. 

Consistency, exchangeability over $A$ in the randomized population, and positivity of treatment assignment are expected to hold for well-defined interventions assessed in (marginally or conditionally) randomized trials. In contrast, positivity of trial participation can be challenging to verify in the data \cite{petersen2012diagnosing} and exchangeability over $S$ is a strong and untestable assumption. Thus, both positivity of trial participation and exchangeability over $S$ need to be examined on the basis of substantive knowledge when extending inferences from a trial to a new target population. 

Now that we have stated the identifiability conditions, we can revisit the relationship of the target population and the trial-eligible population. As we noted in Section \ref{section_designs}, in transportability analyses the target population includes individuals who are not trial-eligible. Conditions (4) and (5) suggest that such transportability analyses are possible only to the extent that some of the variables defining the trial eligibility criteria are not needed to ensure exchangebility over $S$ (e.g., maybe the trial restricted eligibility to particular geographic locations but the investigators believe that conditional on other variables location itself is not associated with the outcome). If all variables used to define the trial eligibility criteria are needed for exchangeability over $S$ then the target population has to be limited to \emph{trial-eligible} non-participants.

\subsection{Identification}

Under the above assumptions, as shown in the Appendix, for each treatment $a$, the potential outcome mean in the target population can be identified using the observed data, 
\begin{equation*}
\E [Y^a | S = 0 ] = \E \big[ \E [ Y | X, S = 1, A = a] \big| S = 0 \big]. 
\end{equation*}
The average treatment effect comparing treatments $a$ and $a^\prime$ in the target population can be identified as 
\begin{equation*}
\E [Y^a - Y^{a^\prime} | S = 0 ] = \E \big[ \E [ Y | X, S = 1, A = a] | S = 0 \big] - \E \big[ \E [ Y | X, S = 1, A = a'] \big| S = 0 \big]. 
\end{equation*}

In the Appendix we show that the average treatment effect in the target population is identifiable under an assumption of conditional exchangeability in measure over $S$, which is weaker than conditional exchangeability in mean over $S$. Under the weaker assumption, we can identify average treatment effects but we cannot identify potential outcome means. Because these means are of inherent scientific and policy interest, in the next section we focus on methods for estimating the functionals $\mu(a) \equiv \E \big[ \E [ Y | X, S = 1, A = a] \big| S = 0 \big]$.

\section{Estimation} \label{section_estimators}

We consider three approaches for estimating $\mu(a)$: (1) \emph{outcome modeling}; (2) \emph{probability of trial participation modeling} followed by weighting; and (3) \emph{doubly robust} approaches that combine outcome and trial participation modeling. Under the consistency assumption, these methods can be thought of as solutions to a missing data problem for the outcome, where the missing data indicator is $I(S = 1, A = a)$. In the main text of the tutorial, we focus on developing intuition about the methods; in the Appendix, we provide some theoretical results. Throughout, we assume that the working models are parametric, as is usually the case in applied work (we consider the use of more flexible methods, such as machine learning, in the Discussion).

\subsection{Outcome modeling}\label{outcome_model}

The first approach extends inferences from trial participants to the population of non-participants by applying an outcome regression model fit among the former to a sample of the latter \cite{robins1986}. In essence, we use data from trial participants to estimate models for the expectation of the outcome and then average the model predictions with respect to the covariate distribution of the non-participants. The outcome model-based estimator (OM) of $\mu(a)$ is
\begin{equation}\label{estimator_OR}
\widehat\mu_{\text{\tiny OM}}(a)   =  \Bigg\{  \sum\limits_{i=1}^{n} (1 - S_i)  \Bigg\}^{-1}    \sum\limits_{i = 1}^{n}  (1 - S_i) g_{a}(X_i; \widehat{\theta})    ,
\end{equation}
where $g_{a}(X; \widehat{\theta})$ is is an estimator for $\E[Y | X, S = 1, A = a]$ based on a parametric model with finite-dimensional parameter $\theta$. We usually estimate separate outcome models in each treatment group of the trial, if the data allow, to account for all possible treatment $\times$ covariate interactions. When the model is correctly specified the estimator is consistent for $\mu(a)$.

\subsection{Trial participation modeling and weighting}\label{sub_section_weighting}

The second approach extends inferences from trial participants to the population of non-participants using a model for the probability of trial participation \cite{cole2010, westreich2017, dahabreh2018generalizing}. In essence, we are treating the randomized trial participants as a sample from the target population with sampling probabilities that depend on baseline covariates and need to be estimated, an idea that connects this approach to survey sampling \cite{Horvitz1952}. Specifically, we estimate $\mu(a)$ using the IO weighting estimator 
\begin{equation}\label{estimator_IOW1}
\widehat\mu_{\text{\tiny IOW1}}(a)  =  \Bigg\{ \sum\limits_{i=1}^{n} (1 - S_i) \Bigg\}^{-1}  \sum\limits_{i = 1}^{n}    \widehat w_a(X_i, S_i, A_i) Y_i,
\end{equation}
with the estimated weights $\widehat w_a(X_i, S_i, A_i) $ defined as $$\widehat w_a(X_i, S_i, A_i) = \dfrac{1 - p(X_i; \widehat\beta)}{ p(X_i; \widehat\beta) e_a(X_i; \widehat\gamma) } \times I(S_i = 1, A_i = a).$$ 
Here, $p(X; \widehat\beta)$ is an estimator for the probability of participation in the trial, $\Pr[S = 1 | X]$, based on a parametric model with finite-dimensional parameter $\beta$, and $e_a(X; \widehat\gamma)$ is an estimator for the probability of being assigned to treatment $a$ among trial participants, $\Pr[A  = a | X, S = 1]$, based on a parametric model with finite-dimensional parameter $\gamma$. The probability of participation in the trial is, in general, unknown and has to be estimated (e.g., by fitting a logistic regression model). In contrast, the probability of treatment in the randomized trial is known (determined by the investigators) and the true values can be used instead of estimated values (estimating the probability may, however, lead to smaller standard errors). We refer to the estimator in (\ref{estimator_IOW1}) as an inverse odds (IO) weighting estimator because $\dfrac{ 1- p(X_i; \widehat\beta) }{ p(X_i; \widehat\beta)}$ is the inverse of the estimated odds of trial participation conditional on baseline covariates.

An alternative IO weighting estimator normalizes the weights to sum to the number of non-randomized individuals,
\begin{equation} \label{estimator_IOW2}
\widehat\mu_{\text{\tiny IOW2}}(a)  =   \Bigg\{   \sum\limits_{i = 1}^{n} \widehat w_a(X_i, S_i, A_i)  \Bigg\}^{-1} \sum\limits_{i = 1}^{n}   \widehat w_a(X_i, S_i, A_i) Y_i .
\end{equation}
The above normalized IO estimator can be obtained using a (saturated) weighted least squares regression of the outcome on treatment, using weights equal to $ \widehat w_a(X_i, S_i, A_i)$; for example, for binary treatment $\widehat\mu_{\text{\tiny IOW2}}(a=0)$ is equal to the intercept of the regression and  $\widehat\mu_{\text{\tiny IOW2}}(a=1)$ is equal to the sum of the intercept and the coefficient of treatment. Because of the normalization by the sum of the weights, this estimator produces estimates that are always within the support of the outcome random variable (e.g., if $Y$ is binary the estimates are between 0 and 1), resulting in better finite-sample performance when the weights are highly variable. Estimators like this one are sometimes referred to as ``ratio estimators'' and have a long history in survey research \cite{hajek1971comment}.

When the model for the probability of participation is correctly specified both IO weighting estimators are consistent for $\mu(a)$. The small difference in the normalization of the IO weights can have a big impact on estimator behavior when weights are highly variable \cite{Robins2007}, because estimator (\ref{estimator_IOW1}) may produce estimates that fall outside the support of the outcome variable, whereas estimator (\ref{estimator_IOW2}) always produces results that fall in the support of the outcome variable.

\subsection{Doubly robust estimators}

In practical applications, background knowledge is typically inadequate to ensure correct specification of the working models for the probability of participation or the expectation of the outcome, and misspecification of these models can lead to estimator inconsistency. We can gain some robustness to misspecification compared to weighting estimators by combining the two models to obtain \emph{doubly robust} estimators that are consistent when either model is correctly specified \cite{robins1994estimation, robins1995semiparametric, Robins2001}. Here, we examine three doubly robust estimators that are easy to implement in standard statistical software. 

The first doubly robust estimator we consider relies on estimating models for the conditional expectation of the outcome, $g_a(X; \theta)$; the probability of trial participation, $p(X; \beta)$; and (optionally) the probability of treatment among trial participants, $e_a(X; \gamma)$. Predicted values from these models are then combined to obtain the estimator  
\begin{equation}\label{estimator_DR1}
\widehat\mu_{\text{\tiny DR1}}(a)  =   \Bigg\{ \sum\limits_{i=1}^{n} (1 - S_i) \Bigg\}^{-1}    \sum\limits_{i = 1}^{n} \Bigg\{\widehat w_a(X_i, S_i, A_i) \Big\{ Y_i - g_a(X_i; \widehat\theta) \Big\}   + (1 - S_i) g_a(X_i; \widehat\theta) \Bigg\}, 
\end{equation}
where $\widehat w_a(X_i, S_i, A_i)$ was defined in the previous section. 

Using the normalized IO weights, an alternative doubly robust estimator is
\begin{equation}\label{estimator_DR2}
	\begin{split}
\widehat\mu_{\text{\tiny DR2}}(a)  & =   \Bigg\{  \sum\limits_{i = 1}^{n} \widehat w_a(X_i, S_i, A_i) \Bigg\}^{-1}    \sum\limits_{i = 1}^{n} \widehat w_a(X_i, S_i, A_i) \Big\{ Y_i - g_a(X_i; \widehat\theta)  \Big\}  + \Bigg\{ \sum\limits_{i=1}^{n} (1 - S_i) \Bigg\}^{-1}  \sum\limits_{i = 1}^{n} (1 - S_i) g_a(X_i; \widehat\theta).
	\end{split}
\end{equation}

A third doubly robust estimator involves fitting a model for the outcome conditional on covariates among trial participants, using a weighted regression with the weights $\widehat w_a(X_i, S_i, A_i)$ as defined above, and then averaging the predicted values, $g_a (X_i; \widehat{\theta}_{\text{\tiny w}})$ over the covariate distribution of non-participants,
\begin{equation}\label{estimator_DR3}
\widehat\mu_{\text{\tiny DR3}}(a)  = \Bigg\{  \sum\limits_{i = 1}^{n} (1 - S_i) \Bigg\}^{-1}   \sum\limits_{i = 1}^{n} (1 - S_i) g_a (X_i; \widehat{\theta}_{\text{\tiny w}}) ,
\end{equation}
where $\widehat{\theta}_{\text{\tiny w}}$ is the vector of estimated parameters from the weighted outcome regression. This estimator is doubly robust when the outcome is modeled with a linear exponential family quasi-likelihood \cite{gourieroux1984} and the canonical link function \cite{Robins2007, Wooldridge2007}.

Doubly robust estimators are consistent for $\mu(a)$ and asymptotically normal when either the model for the probability of participation or the expectation of the outcome is correctly specified \cite{robins1995semiparametric, Tsiatis2007} (see the Appendix for a formal result). More important for applied work is that doubly robust estimators usually produce estimates that are more precise than those of IO weighting estimators and nearly as precise as those of the outcome model-based estimator, while giving investigators two oportunities for valid inference. In some cases, however, when misspecification of the outcome model is combined with highly variable weights doubly robust estimators can perform worse than non-doubly robust estimators that use the same misspecified outcome model \cite{Kang2007, Robins2007}.

\subsection{Inference}

When using parametric working models, all the estimators described above can be viewed as partial M-estimators \cite{Stefanski2002} and it is possible to employ the usual ``sandwich'' approach to obtain their sampling variances (e.g., \cite{Lunceford2004,chen2017variance}). Inference based on the non-parametric bootstrap \cite{efron1994introduction}, however, is easy to obtain and will often be preferred in practice.

\section{Simulation study} \label{section_simulations}

We conducted proof-of-concept simulation studies to examine the finite-sample performance of different estimators for potential outcome means and treatment effects in the target population. In this Section, we describe simulation studies with continuous outcomes in nested trial designs and present results from select scenarios from those simulations. In the Appendix, we report complete results from simulations with continuous and binary outcomes for nested trial designs, as well as the methods and results of additional simulation studies for non-nested trial designs.

\subsection{Data generation}

We run a factorial experiment using 3 trial sample sizes ($n_{\text{\tiny RCT}}$) $\times$ 3 total cohort sample sizes ($n$) $\times$ 2 magnitudes of departure from additive effects in the outcome model $\times$ 2 magnitudes of selection into the trial, resulting in a total of 36 simulation scenarios. 

We considered cohorts of $n =$ 2000, 5000, or 10,000 individuals with the number of randomized individuals  $n_{\text{\tiny RCT}}$ approximately equal to 200, 500, or 1000. Specifically, we generated baseline covariates for cohorts of 2000, 5000, or 10,000 individuals as $X = (1, X_1, X_2, X_3)^\intercal$, and for $j=1,2,3$, $X_j$ followed an independent standard normal distributions. We then simulated selection into the trial using a binary indicator $S$ with $\mbox{logit} \Pr[S = 1 | X] = \beta X$, with $\beta = (\beta_0, \ldots, \beta_3)$. We chose $\beta_0$ using numerical methods \cite{austin2008}, such that for a given cohort sample size $n$ the trial sample size $n_{\text{\tiny RCT}}$ was 200, 500, or 1000. We examined scenarios with $\beta_1 = 0$, indicating no selection on $X_1$, and $\beta_1 = 1$, indicating strong selection; in all scenarios we used $\beta_2 = \beta_3 = 1$, indicating strong selection on $X_2$ and $X_3$.

We generated potential outcomes as $Y^a = \theta^a X + \epsilon^a$. We used $\theta^1 = (1, \theta_1^1, 1, 1)$, with $\theta_1^1 = 2$ or $1$. We used $\theta^0 = ( 0,1,1,1)$ in all scenarios. We generated the errors $\epsilon^a$ using independent standard normal distributions. Note that $\theta_1^1 = 2$ indicates the presence of effect modification on the mean difference scale. For observations ``selected'' into the trial ($S=1$) we generated treatment assignment $A$ as a binomial random variable with parameter 0.5 (i.e., we simulated marginally randomized trials). Last, we generated observed outcomes as $Y = A Y^1  + (1 - A) Y^0$.

For each simulated dataset, we applied the estimators in equations (\ref{estimator_OR}) through (\ref{estimator_DR3}), and also obtained a trial-only estimator of the treatment effect. All working models required for the different estimators were correctly specified, in the sense that the true models were nested within the parametric working models on which the estimators relied. Specifically, outcome models included main effects for all covariates and were fit separately in each arm;  logistic regression models for trial participation and treatment included the main effects of all covariates; all models had intercept terms. We estimated the bias and variance for each estimator over 100,000 runs for each scenario.

\subsection{Simulation results}

Tables \ref{table_simulation_bias} and \ref{table_simulation_variance} summarize simulation results from selected simulation scenarios for continuous, normally distributed outcomes, and linear outcome models in nested trial designs. Additional simulation results are presented in the Appendix, including scenarios with binary outcomes and non-nested trial designs. 

When all models were correctly specified, all estimators were approximately unbiased, even with fairly small trial and target population sample sizes. The outcome-model based estimator had the lowest variance, followed closely by the three doubly robust estimators. The IO weighting estimators had substantially larger variance than all other estimators; that variance, though, became smaller with increasing trial sample sizes. When trial sample size was much smaller than the target sample size, in the presence of strong selection on covariates, estimators that used normalized weights (IOW2, DR2, and DR3 in the Tables) had smaller variance compared to estimators that used unnormalized weights (IOW1 and DR1 in the Tables). As expected, in the presence of selection on a covariate that is also an effect modifier, the trial-only estimator gave different results compared to the estimators in equations (\ref{estimator_OR}) through (\ref{estimator_DR3}). Specifically, the trial-only estimator is biased for $\E [Y^1 - Y^0 | S = 0 ]$ when selection into the trial depends on the effect modifier, but is, of course, unbiased for $\E [Y^1 - Y^0 | S = 1]$ under very general conditions. 

\emph{A note about model misspecification in our simulation studies:} Even though the simulations did not explicitly examine misspecified models, they nevertheless provide a good illustration of the double robustness property! This counter-intuitive claim is justified because the IO weighting estimators can be viewed as special versions of the doubly robust estimators with grossly misspecified models of the expectation of the outcome; specifically, models that produce predictions $g_a(X; \widehat \theta)$ identically equal to 0. Similarly, the outcome model-based estimators can be viewed as special versions of the doubly robust estimators with grossly misspecified models for the probability of trial participation; specifically, models that produce probabilities of trial participation $p(X; \widehat \beta)$  identically equal to 1, so that the weights $\widehat w_a(X, S, A)$ are identically equal to 0. Because our IO weighting and outcome regression estimators (with correctly specified probability of participation and outcome models, respectively) are approximately unbiased,  viewing these estimators as versions of doubly robust estimators (with grossly misspecified outcome or probability of participation models, respectively), verifies the double robustness property.

\begin{table}[ht!]
\centering
\caption{Selected results from the simulation study: bias of estimators across sample sizes, under strong selection on the effect modifier ($\beta_1 = 1$), strong effect modification ($\theta_1^1 - \theta_1^0 = 1$).}
\label{table_simulation_bias}
\begin{tabular}{ccccccccc}
\toprule
$n_{\text{\tiny RCT}}$ & $n$ &  \textbf{Trial} & \textbf{IOW1} & \textbf{IOW2} & \textbf{OM}  &  \textbf{DR1} & \textbf{DR2} & \textbf{DR3} \\ \midrule
200 & 	2000 & 	0.7656 & 	0.0018 & 	0.0593 & 	0.0000 & 	0.0000 & 	0.0003 & 	0.0006 \\ 
200 & 	5000 & 	0.8401 & 	0.0080 & 	0.0710 & 	-0.0007 & 	0.0002 & 	-0.0002 & 	0.0000 \\
200 & 	10000 & 	0.8855 & 	0.0029 & 	0.0753 & 	-0.0002 & 	-0.0019 & 	-0.0015 & 	-0.0012 \\
500 & 	2000 & 	0.6892 & 	0.0036 & 	0.0235 & 	0.0001 & 	0.0016 & 	0.0013 & 	0.0013 \\ 
500 & 	5000 & 	0.7654 & 	-0.0003 & 	0.0306 & 	-0.0001 & 	0.0004 & 	0.0002 & 	-0.0004 \\ 
500 & 	10000 & 	0.8233 & 	0.0033 & 	0.0339 & 	-0.0012 & 	-0.0023 & 	-0.0021 & 	-0.0020 \\ \midrule 
1000 & 	2000 & 	0.6576 & 	-0.0004 & 	0.0081 & 	0.0000 & 	-0.0004 & 	-0.0004 & 	-0.0003 \\ 
1000 & 	5000 & 	0.7061 & 	0.0033 & 	0.0157 & 	-0.0003 & 	-0.0009 & 	-0.0009 & 	-0.0005 \\
1000 & 	10000 & 	0.7653 & 	-0.0010 & 	0.0160 & 	0.0002 & 	0.0002 & 	0.0002 & 	0.0002 \\ \bottomrule
\end{tabular}
\caption*{Trial = trial-only, unadjusted estimator; OM = outcome model-based estimator; IOW1 = inverse odds weighting estimator without normalized weights; IOW2 = inverse odds estimator with normalized weights; DR1 = doubly robust estimator without normalized weights; DR2 = doubly robust estimator with normalized weights; DR3 = doubly robust estimator based on weighted multi-variable regression. Additional simulation results are presented in the Appendix.}
\end{table}

\begin{table}[ht!]
\centering
\caption{Selected results from the simulation study: variance of estimators across sample sizes, under strong selection on the effect modifier ($\beta_1 = 1$) and strong effect modification ($\theta_1^1 - \theta_1^0 = 1$).}
\label{table_simulation_variance}
\begin{tabular}{ccccccccc}
\toprule
$n_{\text{\tiny RCT}}$ & $n$ &  \textbf{Trial} & \textbf{IOW1} & \textbf{IOW2} & \textbf{OM}  &  \textbf{DR1} & \textbf{DR2} & \textbf{DR3} \\ \midrule
200 & 	2000 & 	0.0843 & 	4.4507 & 	1.1256 & 	0.0762 & 	0.2562 & 	0.1501 & 	0.1582 \\
200 & 	5000 & 	0.0905 & 	6.9650 & 	1.2765 & 	0.0819 & 	0.3353 & 	0.1669 & 	0.1791 \\
200 & 	10000 & 	0.0939 & 	6.1245 & 	1.3539 & 	0.0841 & 	0.3321 & 	0.1753 & 	0.1921 \\  
500 & 	2000 & 	0.0324 & 	1.1901 & 	0.5278 & 	0.0268 & 	0.0808 & 	0.0610 & 	0.0558 \\ 
500 & 	5000 & 	0.0336 & 	1.5804 & 	0.6760 & 	0.0294 & 	0.1059 & 	0.0718 & 	0.0662 \\ 
500 & 	10000 & 	0.0351 & 	1.4593 & 	0.7505 & 	0.0310 & 	0.1113 & 	0.0783 & 	0.0727 \\  
1000 & 	2000 & 	0.0166 & 	0.4580 & 	0.2456 & 	0.0127 & 	0.0333 & 	0.0290 & 	0.0258 \\ 
1000 & 	5000 & 	0.0162 & 	0.6061 & 	0.3558 & 	0.0137 & 	0.0423 & 	0.0346 & 	0.0305 \\  
1000 & 	10000 & 	0.0168 & 	0.6964 & 	0.4291 & 	0.0146 & 	0.0503 & 	0.0399 & 	0.0352 \\ \bottomrule
\end{tabular}
\caption*{Trial = trial-only, unadjusted estimator; OM = outcome model-based estimator; IOW1 = inverse odds weighting estimator without normalized weights; IOW2 = inverse odds weighting estimator with normalized weights; DR1 = doubly robust estimator without normalized weights; DR2 = doubly robust estimator with normalized weights; DR3 = doubly robust estimator based on weighted multivariable regression. Additional simulation results are presented in the Appendix.}
\end{table}

\subsection{Code to implement the methods}

In the Appendix, we provide \texttt{R} \cite{RCT2015} code implementing the methods compared in the simulation study. Specifically, we provide a collection of basic stand-alone functions, one for each estimator in equations (\ref{estimator_OR}) through (\ref{estimator_DR3}), using parametric working models estimated by standard maximum likelihood methods. Readers can modify the functions to incorporate alternative estimation approaches and to obtain bootstrap-based inference. To allow inference with sandwich standard errors, we also provide an implementation of the estimators using the \texttt{R} package \texttt{geex} \cite{Saul2017}. Last, we provide \texttt{Stata} code to reproduce the simulation study.

\section{Analyses in the Coronary Artery Surgery Study} \label{section_example}

The Coronary Artery Surgery Study (CASS) included a randomized trial nested within a cohort study (i.e., could be viewed as a nested trial design in the terminology of Section \ref{section_designs}), comparing coronary artery surgery plus medical therapy (henceforth, ``surgery'') versus medical therapy alone for patients with chronic coronary artery disease. Of the 2099 eligible patients, 780 consented to randomization and 1319 declined. We excluded six patients for consistency with prior CASS analyses \cite{chaitman1990, olschewski1992} and in accordance with CASS data release notes; in total, we used data from a total of 2093 patients. Details about the design of the CASS are available elsewhere \cite{william1983, investigators1984}. Here, we focus on estimating the survival probability and treatment effects among eligible non-participants and comparing them against estimates obtained among trial participants.

We implemented the methods described in Section \ref{section_estimators} to estimate the 10-year risk (cumulative incidence proportion) of death from any cause in the surgery and medical therapy groups, the risk difference, and the relative risk for the population of eligible patients who did not consent to randomization. Risks are reasonable measures of incidence in CASS because no patients were censored during the first 10 years of follow-up. The working models for the outcome, the probability of participation in the trial, and the probability of treatment were logistic regression models with the following covariates: age, severity of angina, history of myocardial infarction, percent obstruction of the proximal left anterior descending artery, left ventricular wall motion score, number of diseased vessels, and ejection fraction. We selected variables for inclusion in the models based on a previous analysis of the CASS data \cite{olschewski1992}; age and ejection fraction were modeled using restricted cubic splines with 5 knots \cite{Harrell2015}. We used bootstrap re-sampling with 10,000 samples to obtain percentile-based 95\% confidence intervals. 

\begin{figure}[bt]
	\centering
	\caption{Kernel densities for the estimated probability of trial participation for trial participants (left panel, solid line) and non-participants (left panel, dashed line), and the estimated weights for trial participants (right panel). The weights for trial participants are equal to the inverse of the estimated odds of trial participation times the inverse of the estimated probability of receiving the treatment actually received, as defined in Section \ref{sub_section_weighting}.}
	\includegraphics[width=12cm]{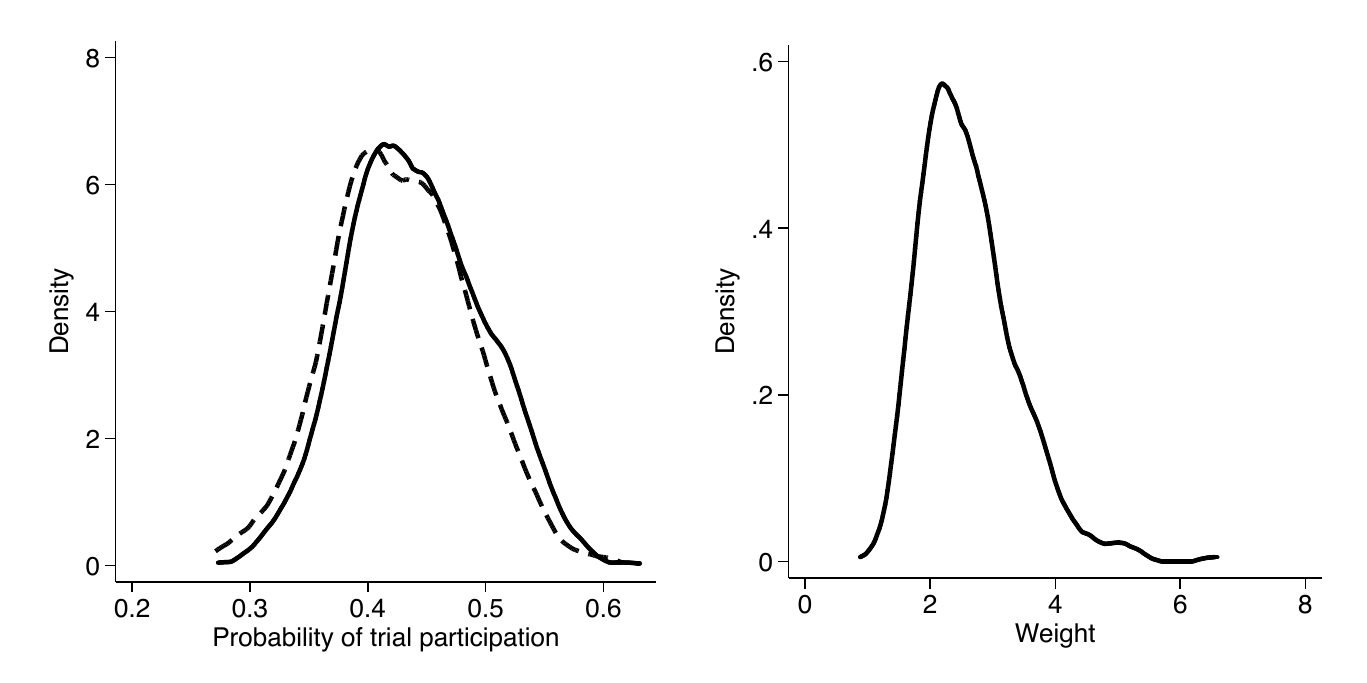}
	\label{fig:densities}
\end{figure}

Of the 2093 patients in the CASS dataset, 1686 had complete data on all baseline covariates (731 randomized, 368 to surgery and 363 to medical therapy; 955 non-randomized, 430 receiving surgery and 525 medical therapy). For simplicity, we only report analyses restricted to patients with complete data. Table \ref{Table_4_cass_baseline} summarizes baseline covariates in trial participants and non-participants (by treatment group). Figure \ref{fig:densities} presents the kernel density of the estimated probability of trial participation for trial participants and non-participants and a kernel density of the estimated weights for trial participants. The sample proportion of non-participants divided by the sample average of the IO of trial participation among trial participants was approximately 1.001. We present a comparison of the post-weighting covariate distribution in trial participants against the covariate distribution in non-participants in Section E of the Appendix.

Estimates of the 10-year risk (by treatment group), risk difference, and risk ratio are shown in Table \ref{table_CASS}. The outcome model-based estimator (OM), the IO weighting estimators (IOW1 and IOW2), and the doubly robust estimators (DR1, DR2, and DR3) produced similar results, suggesting that findings are not driven by model specification decisions \cite{Robins2001}. Furthermore, the results from these estimators were similar to those from the trial-only analyses, reflecting the similarity of covariate distributions between randomized and non-randomized individuals. In general, differences will be more striking when the covariate distributions are more different and in the presence of strong effect modification by measured variables. Nevertheless, conducting the analyses we describe in this tutorial is necessary to obtain proper estimates of uncertainty for the potential outcome means or treatment effects in the target population.

\begin{table}[ht!]
\centering
\caption{Baseline covariates in the CASS (August 1, 1975 to Dec 31, 1989).}
\label{Table_4_cass_baseline}
\begin{tabular}{llcccc}
\toprule
\multirow{2}{*}{\textbf{Variable}} & \multirow{2}{*}{\textbf{Levels}} & \multicolumn{2}{c}{\textbf{Non-randomized group}}                                                                                           & \multicolumn{2}{c}{\textbf{Randomized group}}                                                                                               \\
                                   &                                  & \textbf{\begin{tabular}[c]{@{}c@{}}Surgery\\ (430)\end{tabular}} & \textbf{\begin{tabular}[c]{@{}c@{}}Medical therapy\\ (525)\end{tabular}} & \textbf{\begin{tabular}[c]{@{}c@{}}Surgery\\ (368)\end{tabular}} & \textbf{\begin{tabular}[c]{@{}c@{}}Medical therapy\\ (363)\end{tabular}} \\ \midrule
Age, years                         &                                  & 51.3 (7.7)                                                       & 50.6 (7.8)                                                               & 51.4 (7.2)                                                       & 50.9 (7.4)                                                               \\
Angina                             & None                             & 70 (16.3\%)                                                      & 125 (23.8\%)                                                             & 83 (22.6\%)                                                      & 81 (22.3\%)                                                              \\
                                   & Present                          & 360 (83.7\%)                                                     & 400 (76.2\%)                                                             & 285 (77.4\%)                                                     & 282 (77.7\%)                                                             \\
History of MI                      & No                               & 194 (45.1\%)                                                     & 212 (40.4\%)                                                             & 159 (43.2\%)                                                     & 135 (37.2\%)                                                             \\
                                   & Yes                              & 236 (54.9\%)                                                     & 313 (59.6\%)                                                             & 209 (56.8\%)                                                     & 228 (62.8\%)                                                             \\
LAD \% obstruction                 &                                  & 48.2 (39.8)                                                      & 31.7 (36.2)                                                              & 36.4 (38.0)                                                      & 34.9 (37.0)                                                              \\
Left ventricular score             &                                  & 7.1 (2.7)                                                        & 7.1 (2.7)                                                                & 7.4 (2.9)                                                        & 7.3 (2.8)                                                                \\
Diseased vessels                   & 0  & 113 (26.3\%)       & 234 (44.6\%)                                                             & 146 (39.7\%)                                                     & 133 (36.6\%)                                                             \\
                                   & $\geq$ 1                & 317 (73.7\%)                                                     & 291 (55.4\%)                                                             & 222 (60.3\%)                                                     & 230 (63.4\%)                                                             \\
Ejection fraction, \%              &                                  & 60.2 (12.0)                                                      & 60.1 (12.5)                                                              & 60.9 (13.0)                                                      & 59.8 (12.8)    \\      \bottomrule
\end{tabular}
\caption*{Results presented as mean (SD) for continuous variables and count (\%) for discrete variables. \\
CASS = Coronary Artery Surgery Study; LAD = left anterior descending coronary artery; MI = myocardial infarction; SD = standard deviation.}
\end{table}

\begin{table}[ht!]
\centering
\caption{Estimated 10-year mortality risk, risk  difference, and risk ratio for surgery vs. medical therapy among trial participants and non-participants in the Coronary Artery Surgery Study (August 1, 1975 to Dec 31, 1989).}
\label{table_CASS}
\begin{tabular}{ccccc}
\toprule
\textbf{Estimator} &  \begin{tabular}[c]{@{}c@{}}\textbf{Survival probability} \\ \textbf{(Surgery)}\end{tabular} & \begin{tabular}[c]{@{}c@{}}\textbf{Survival probability} \\ \textbf{(Medical)}\end{tabular} & \textbf{Risk difference}  & \textbf{Risk ratio}  \\ \midrule
Trial-only & 17.4\% (13.6\%, 21.4\%) & 20.4\% (16.3\%, 24.6\%) & -3.0\% (-8.7\%, 2.7\%) & 0.85 (0.62, 1.15) \\
OM         & 18.9\% (13.9\%, 22.7\%) & 20.1\% (15.9\%, 24.5\%) & -1.3\% (-7.9\%, 4.2\%) & 0.94 (0.65, 1.24) \\
IOW1       & 18.2\% (13.9\%, 22.7\%) & 20.1\% (15.9\%, 24.4\%) & -1.9\% (-7.8\%, 4.2\%) & 0.91 (0.66, 1.24) \\
IOW2       & 18.2\% (14.6\%, 23.5\%) & 20.1\% (16.0\%, 24.4\%) & -1.9\% (-7.2\%, 4.8\%) & 0.91 (0.69, 1.28) \\
DR1        & 18.7\% (14.5\%, 23.3\%) & 20.1\% (16.0\%, 24.4\%) & -1.4\% (-7.3\%, 4.7\%) & 0.93 (0.68, 1.27) \\
DR2        & 18.7\% (14.5\%, 23.3\%) & 20.1\% (16.0\%, 24.4\%) & -1.4\% (-7.3\%, 4.7\%) & 0.93 (0.68, 1.27) \\
DR3        & 18.7\% (14.4\%, 23.2\%) & 20.0\% (15.9\%, 24.3\%) & -1.4\% (-7.3\%, 4.6\%) & 0.93 (0.68, 1.27) \\ \bottomrule
\end{tabular}
\caption*{Numbers in parentheses are 95\% non-parametric percentile bootstrap confidence intervals based on 10,000 bootstrap samples. Trial = trial-only, unadjusted estimator; OM = outcome model-based estimator; IOW1 = inverse odds weighting estimator without normalized weights; IOW2 = inverse odds weighting estimator with normalized weights; DR1 = doubly robust estimator without normalized weights; DR2 = doubly robust estimator with normalized weights; DR3 = doubly robust estimator based on weighted multi-variable regression; Surgery = coronary artery bypass grafting surgery plus medical therapy; Medical = medical therapy.}
\end{table}

\section{Practical issues}

We now discuss practical issues related to lack of exchangeability over $S$, covariate selection, other aspects of model specification, and positivity violations and highly variable weights, all of which can affect applied analyses.

\subsection{Violations of the exchangeability assumption}

All methods described in this tutorial depend critically on the assumption of conditional mean exchangeability over $S$. When the assumption does not hold, the results produced by the various estimators described in this paper do not have a causal interpretation. Because the mean exchangeability assumption is not testable, it is reasonable to explore how conclusions would change under different degrees of assumption violations. Depending on the available data and background knowledge, various sensitivity analysis methods for missing data or unmeasured confounding can be modified for use in analyses extending inferences to a target population (e.g., \cite{nguyen2017sensitivity, nguyen2018sensitivity, dahabreh2019sensitivitybiascor}). In a companion paper, we have proposed easy to implement estimators for sensitivity analysis that are based on the estimators described in Section \ref{section_estimators} and do not require extensive background knowledge about the distribution of unknown or unmeasured variables that are the source of violations of the exchangeability condition \cite{dahabreh2019sensitivitybiascor}.

\subsection{Covariate selection} 

Throughout, we have used $X$ to signify baseline covariates measured both among randomized trial participants and the sample from the target population. In principle, investigators can use any subset of the available covariates that satisfies the assumption of conditional exchangeability over $S$. When investigators are interested in estimating the potential outcome mean under each treatment (not only the average treatment effect), outcome predictors that are also associated with trial participation should be included in models for the outcome and the probability of participation (or both, when using doubly robust estimators). Including outcome predictors that are not associated with trial participation in models for the expectation of the outcome will often decrease the variance of the outcome model-based and doubly robust estimators; including strong predictors of trial participation that are not associated with the outcome in regressions for the probability of participation will increase the variance of the IO weighting and doubly robust estimators without improving the ability to extend inferences to the target population. When investigators are primarily interested in the average treatment effect (instead of the potential outcome mean under each treatment), only effect modifiers (on the mean difference scale) need to be modeled \cite{zhang2015, omuircheartaigh2014}. Because background knowledge about effect modification is typically limited, even when interest is centered on treatment effect estimation, it is probably best to include as many outcome predictors as possible in regression models for the expectation of the outcome or the probability of trial participation. We followed this strategy in our CASS re-analysis: we selected covariates for ``adjustment'' based on prior work on outcome modeling and used the same covariates when modeling trial participation, the outcome, and treatment in the trial.

\subsection{Other aspects of model specification} 

Models for the expectation of the outcome and the probability of trial participation need to approximate the corresponding ``true'' conditional expectation/probability functions. Covariate balance assessments \cite{rubin2001using} comparing the post-weighting covariate distribution in trial participants against the (unweighted) covariate distribution in non-participants can provide some guidance for the choice of functional form. When considering parametric models, the methods described in \cite{waernbaum2017model} can be easily modified to derive the asymptotic bias induced by model misspecification for the estimators described in our paper.

To mitigate model misspecification, investigators can use flexible parametric models (e.g., by including splines for continuous variables or interactions between regressors) or semi- and non-parametric models (e.g., machine learning). When using very flexible models for the outcome or the probability of participation, investigators need to consider the rate of convergence of the working model estimators and its impact on the behavior of potential outcome mean or treatment effect estimators. A detailed discussion of this subtle issue is beyond the scope of the present tutorial (see, \cite{chernozhukov2018double} for a detailed treatment), but in general doubly robust estimators should be the approach of choice in this context, because the estimators remain $\sqrt n$-consistent even when using estimators for the conditional expectation of the outcome or the probability of participation that converge at slower than $\sqrt n$-rate \cite{robins1997toward, chernozhukov2018double}. The doubly robust estimators of $\mu(a)$ and the machine learning methods (for estimating the conditional expectation of the outcome or the conditional probability of trial participation) can be combined with cross-fitting methods (a sample splitting approach that reduces overfitting bias) to achieve good behavior under weak assumptions about the underlying models and the properties of the machine learning methods \cite{chernozhukov2018double}. Also in the spirit of flexible model specification, for outcome model-based or doubly robust estimators, we recommend fitting separate regression models for the outcome in each treatment group in the trial, as we did in the CASS re-analysis (equivalent to fitting a single regression model that includes all possible treatment-covariate interactions).

More broadly, model specification for extending inferences from a trial to a target population involves trading off bias and variance using informal \cite{cole2008constructing} or formal methods (e.g., \cite{brookhart2006semiparametric}). When background knowledge suggests that a large number of covariates need to be modeled, formal model specification search methods can be particularly helpful. In our experience, especially when using composite datasets, the variables measured both among trial participants and the sample of the target population are often few and model specification is not a pressing concern (of course, such cases raise concerns about violations of the identifiability conditions and necessitate sensitivity analyses). When richer data are available (e.g., when trials are nested in cohorts of eligible individuals \cite{dahabreh2018generalizing}), we may need to employ more sophisticated strategies for model specification search (e.g., \cite{vansteelandt2012model} provides an overview in the context of causal inference for observational studies, but the same principles apply here).

\subsection{Positivity violations and highly variable weights} To prevent structural violations of the positivity of trial participation assumption, investigators should ensure that the sample from the target population meets the trial eligibility criteria. For example, if the trial restricted enrollment to patients under 85 years of age, it is prudent to apply the same restriction in the sample of patients from the target population. When positivity is violated, IO weighting estimators are inconsistent, whereas outcome model-based and doubly robust estimators rely heavily on the specification of the outcome model (to extrapolate from participants to non-participants) \cite{petersen2012diagnosing}. Empirical (finite-sample) violations of positivity can arise due to chance, particularly when the trial sample size is small or when adjustment for a high-dimensional covariate vector is needed. Empirical violations of positivity increase bias and variance in a way that depends on the particular estimator being used, model specification, and the underlying data generating mechanism. 

It is always a good idea to examine the distribution of the estimated probabilities of trial participation, even when using the outcome model-based estimator, because estimated probabilities of trial participation near zero are warning signs for possible positivity violations. Inspection of the estimated probabilities of trial participation can be combined with diagnostics for positivity violations \cite{petersen2012diagnosing}. It is also useful to inspect the distribution of the weights that are used for the IO weighting and doubly robust estimators. By inspecting the distribution of the IO weights, investigators can identify extreme values and visually assess the spread of the weight distribution. A useful diagnostic is that the sample proportion of non-participants divided by the sample average of the estimated IO of trial participation among trial participants, should be approximately equal to one; or, in symbols $$ \Bigg\{ \sum\limits_{i=1}^n ( 1 - S_i) \Bigg\}^{-1} \sum\limits_{i=1}^n S_i \dfrac{1 - p(X_i; \widehat\beta)}{p(X_i; \widehat\beta)} \approx 1. $$ Values different from 1, suggest positivity violations or model misspecification. The rationale for the diagnostic is provided by the identity $  \E \left[ S \dfrac{  \Pr[S = 0 | X] }{ \Pr[S = 1 | X]} \right] = \Pr[S = 0] . $ 

In applied analyses, we have found that problems with extreme weights can often be addressed by making sensible modeling choices \cite{cole2008constructing} and ensuring that the sample of non-participants is properly selected to avoid violations of the positivity of trial participation assumption. Trimming or truncation of extreme weights may also help, but these strategies shift the causal estimand, which is often undesirable.

In our CASS re-analyses, the estimated probabilities of trial participation were far from zero. Their distribution was similar among trial participants and non-participants (as shown in Figure \ref{fig:densities}), reflecting the fairly similar observed covariate distribution in trial participants and non-participants and the absence of strong selection into the trial (at least based on available covariates, as shown in Table \ref{Table_4_cass_baseline}). As noted, the sample proportion of non-participants divided by the sample average of the IO of trial participation among trial participants was approximately 1, providing some reassurance that gross violations of positivity were absent. After weighting, the covariates included in the model for the probability of trial participation were well-balanced (Appendix Section E).

\section{Discussion} \label{section_discussion}

In this tutorial, we reviewed methods for extending causal inferences about time-fixed treatments from a randomized trial to a target population of non-participants using baseline covariate data from randomized participants and a sample from the target population, but treatment and outcome data only from the randomized participants. We considered estimation approaches that rely on modeling the probability of trial participation, the expectation of the outcome, or both, and can be implemented easily in all popular statistical software packages. 

A major challenge in applying any of the methods discussed in this tutorial is the need to collect adequate covariate information, both from trial participants and non-participants, for the assumption of conditional exchangeability in mean over $S$ to hold. Because this assumption is not testable using the observed data, we have to rely on background substantive knowledge to assess its plausibility. Reasoning about the assumption can be facilitated using directed acyclic graphs, including recent graphical identification algorithms for assessing generalizability/transportability \cite{Bareinboim2013, Pearl2014, bareinboim2012transportability}. Because background knowledge is typically incomplete, it is often necessary to conduct sensitivity analyses, to examine how violations of the exchangeability assumption influence study results \cite{rotnitzky1998semiparametric, Robins2000c}.

Methods related to those discussed in this tutorial have been discussed in a number of recent publications \cite{cole2010, kaizar2011, omuircheartaigh2014, tipton2012, tipton2014, hartman2013, zhang2015, buchanan2018generalizing, rudolph2017, westreich2017}. With few exceptions -- such as the careful asymptotic study of an estimator closely related to DR1 in \cite{zhang2015}, or the targeted maximum likelihood estimators in \cite{rudolph2017} -- prior work has focused on weighting \cite{cole2010,hartman2013,westreich2017} or stratification-based methods \cite{omuircheartaigh2014,tipton2012,tipton2014} that only rely on the probability of trial participation. Theoretical arguments, our simulation results, and practical experience suggest that methods that combine modeling the probability of trial participation with modeling the expectation of the outcome are most promising for applied work for two reasons: first, the double robustness property in effect gives investigators two opportunities for approximately correct inference \cite{Robins2001}; second, doubly robust estimators often produce estimates that are more precise than those from methods that exclusively rely on modeling the probability of trial participation, even when the outcome model is misspecified \cite{Robins2001, Robins2007, Bang2005, Tsiatis2007}. In our proof-of-concept simulation studies, which used correctly specified parametric working models with few covariates, all estimators performed reasonably well in terms of bias. Interestingly, the two IO weighting estimators had very different finite-sample performance in the presence of strong selection. Based on this observation, we recommend avoiding weighted estimators that do not use normalized weights.

When data are available on numerous baseline covariates, many of which are continuous, correct specification of parametric models for the probability of trial participation or expectation of the outcome will be impossible. Future research should address estimation using more flexible models (e.g., non-parametric or semi-parametric regression, machine learning) to mitigate model misspecification \cite{chernozhukov2018double}. Further research is also needed to study the behavior of different estimators under misspecification and to develop alternatives that are more robust to misspecification of the outcome model (e.g., along the lines suggested in \cite{cao2009improving,vermeulen2015bias}). Lastly, throughout this tutorial, we have assumed perfect adherence to treatment in the randomized trial, no missing outcome data, and no measurement error. In practice, adherence is often imperfect, outcomes are missing (e.g., due to right censoring in failure-time analyses), and measurement error is a concern (e.g., differential measurement error in effect modifiers when using composite datasets). Established methods to address these issues in the trial data can be combined with the methods described in this tutorial in a modular fashion. For example, adjustment for imperfect adherence via inverse probability of treatment weighting can be combined with the methods reviewed in this tutorial \cite{dahabreh2019identification}. Future work should assess the properties of such combined procedures and evaluate them in practical applications.

\section*{acknowledgements}
The authors thank Drs. Nina Joyce (Brown University) and John Wong (Tufts Medical Center) for helpful comments on earlier versions of the manuscript. 

This work was supported in part through Patient-Centered Outcomes Research Institute (PCORI) Methods Research Awards ME-1306-03758 and ME-1502-27794 to I.J. Dahabreh, and ME-1503-28119 to M.A. Hern{\'a}n. All statements in this paper, including its findings and conclusions, are solely those of the authors and do not necessarily represent the views of the PCORI, its Board of Governors, or the Methodology Committee.

The data analyses in our paper used CASS research materials obtained from the National Heart, Lung, and Blood Institute (NHLBI) Biologic Specimen and Data Repository Information Coordinating Center. This paper does not necessarily reflect the opinions or views of the CASS or the NHLBI.

\section*{conflict of interest}
The authors have no conflicts of interest to report.

\bibliography{Transporting_the_results_of_experiments_comp}

\end{document}